\documentclass{iaus}
\usepackage{graphics}

 \checkfont{eurm10}
  \iffontfound
    \IfFileExists{upmath.sty}
      {\typeout{^^JFound AMS Euler Roman fonts on the system,
                   using the 'upmath' package.^^J}%
       \usepackage{upmath}}
      {\typeout{^^JFound AMS Euler Roman fonts on the system, but you
                   dont seem to have the}%
       \typeout{'upmath' package installed. iaus.cls can take advantage
                 of these fonts,^^Jif you use 'upmath' package.^^J}%
      }
  \else
  \fi

  \checkfont{msam10}
  \iffontfound
    \IfFileExists{amssymb.sty}
      {\typeout{^^JFound AMS Symbol fonts on the system, using the
                'amssymb' package.^^J}%
       \usepackage{amssymb}%

      }{}
  \fi

  \IfFileExists{amsbsy.sty}
    {\typeout{^^JFound the 'amsbsy' package on the system, using it.^^J}%
     \usepackage{amsbsy}}
    {\providecommand\boldsymbol[1]{\mbox{\boldmath $##1$}}}

\newsavebox{\astrutbox}
\sbox{\astrutbox}{\rule[-5pt]{0pt}{20pt}}

\title{Mapping the Cosmic Web with the Sloan Digital Sky Survey}

\author[Vogeley et al.]{Michael S. Vogeley, Fiona Hoyle, Randall R. Rojas, \& David M. Goldberg}

\affiliation{Department of Physics, Drexel University, 
Philadelphia, PA 19104, email: vogeley@drexel.edu}

\pubyear{2004}
\volume{195}
\pagerange{1--8}
\date{?? and in revised form ??}
\jname{Outskirts of Galaxy Clusters: intense life in the suburbs}
\editors{A. Diaferio, ed.}
\begin{document}

\maketitle

\begin{abstract}
  Wide-angle, moderately deep redshift surveys such as that conducted
  as part of the Sloan Digital Sky Survey (SDSS) allow study of the
  relationship between the structural elements of the large-scale
  distribution of galaxies -- including groups, cluster,
  superclusters, and voids -- and the dependence of galaxy formation
  and evolution on these enviroments. We present a progress report on
  mapping efforts with the SDSS and discuss recently constructed
  catalogs of clusters, voids, and void galaxies, and evidence for a
  $420h^{-1}$Mpc supercluster or ``Great Wall.'' Analysis of
  multi-band photometry and moderate-resolution spectroscopy from the
  SDSS reveals environmental dependence of the star formation history
  of galaxies that extends over more than a factor of 100 in density,
  from clusters all the way to the deep interiors of voids. On
  average, galaxies in the rarified environments of voids exhibit
  bluer colors, higher specific star formation rates, lower dust
  content, and more disk-like morphology than objects in denser
  regions. This trend persists in comparisons of samples in low vs.
  high-density regions with similar luminosity and morphology, thus
  this dependence is not simply an extension of the morphology-density
  relation. Large-scale modulation of the halo mass function
  and the temperature of the intergalactic medium might explain this
  dependence of galaxy evolution on the large-scale environment.
\end{abstract}

\firstsection 

\section{Introduction: Cities, Suburbs, and Countryside of the Universe}

The varied geography and demography of human civilization provides a
colorful analogy for the dependence of galaxy formation and evolution
on the large-scale environment of the universe.  The dense population
of cities promotes a frenetic lifestyle and facilitates many fleeting
interactions. Only the very richest get richer (cD's?), while many, drawn to
the bright lights, are quickly stripped of all that they arrived with.
At the other extreme lie the rural stretches, where interactions are
few but which tend to be tightly-bound when they occur.  Without the
competition of others, a few can live surprisingly well
off the steady, albeit sometimes meager, offerings of the land.
This meeting has been called to study the suburbs of the universe,
including the outskirts of clusters. ``Intense life in the suburbs''
results from living where interactions are frequent, but lasting
enough to form families (groups?) and the city's pull is always
felt. This environment is of particular interest because this is the
lowest density in which halos with multiple galaxies are common.

Galaxy clusters, groups, and filaments surround the voids in a cosmic
web that large galaxy redshift surveys reveal to be a ubiquitous
pattern in the universe. Below we report on progress in mapping the
nearby universe using the SDSS (\S \ref{sec:survey}) and describe results of
objective identification of individual structures in the survey. We then discuss analyses of photometric and spectroscopic
data for galaxy samples that are now large enough to statistically
reveal dependence of galaxy evolution on environment from clusters all
the way to voids (\S \ref{sec:environ}). We conclude by reviewing physical mechanisms that may explain this dependence (\S \ref{sec:discuss}).

\section{Large-Scale Structures in the SDSS}
\label{sec:survey}

To study structure formation in this wide range of environments
requires large statisically-controlled samples of galaxies over a wide
area of sky. The SDSS was designed to accomplish this goal. It employs
a special-purpose 2.5m telescope with a 3-degree field of view, a
mosaic CCD camera, and dual fiber-fed spectrographs, to obtain
five-band digital photometry and moderate-resolution spectroscopy over
essentially the full range of optical wavelengths. The
completed survey will cover approximately $10^{4}$ square degrees.
York et al. (2000) provides an overview of the SDSS and
Stoughton et al. (2002) describes the early data release (EDR) and
details about the photometric and spectroscopic measurements made from
the data.
Technical articles providing details of the SDSS
include descriptions of the photometric camera (Gunn 1998),
photometric analysis (Lupton et al. 2004), the photometric system
(Fukugita et al. 1996; Smith et al. 2002), the photometric monitor
(Hogg et al. 2001), astrometric calibration (Pier et al. 2003),
and spectroscopic tiling (Blanton et
al. 2003).  A thorough analysis of possible systematic uncertainties
in the galaxy samples is described in Scranton et al. (2002).

\subsection{SDSS Galaxy Redshift Surveys}

The SDSS is now the largest galaxy redshift survey to date. When
completed, it will include $10^6$ galaxies over $10^4$ deg$^2$ of sky.
The second major public release of SDSS data (DR2: Abazajian et al. 2004; {\tt www.sdss.org/dr2} and {\tt skyserver.sdss.org}), in March 2004, includes 3324 deg$^2$ of imaging (88 million unique objects) and spectroscopy over 2627 deg$^2$ (including 260,490 galaxies). 

The main spectroscopic galaxy sample of the SDSS (Strauss et al. 2002) includes
objects selected to have Petrosian magnitudes of $r<17.77$ after
correction for Galactic extinction. The median depth of
this sample is approximately $z=0.1$.
Analyses of
these data include measurement of the power spectrum, correlation
function, luminosity function, topology, galaxy properties,
etc. (among 389 SDSS papers so far from collaboration participants; see {\tt www.sdss.org/publications/}).

To obtain a deeper tracer of structure in the universe, a
spectroscopic target sample of luminous red galaxies (LRG's),
is selected by color and has $r<19.5$ (Eisenstein et al. 2001). 
These criteria yield a nearly volume-limited sample out to $z=0.45$.
The LRG selection criteria
ensure that this sample includes all brightest cluster galaxies, as
well as slightly less luminous objects in clusters and groups. The
LRG's are an efficient tracer of dense regions in the universe and
have selection criteria that are quite simple compared to, e.g.,
cluster or group-finding algorithms. The ease of measuring redshifts
to such galaxies (they have strong absorption lines) makes them an
excellent choice for deep mapping of the universe.

\subsection{An Even Greater Wall}

A novel approach to plotting the distribution of objects over large
ranges of cosmic time and distance (Gott et al. 2003) clearly shows
the largest structure seen to date, a ``Sloan Great Wall'' that
extends over roughly $420h^{-1}$Mpc. This supercluster-scale structure
may be compared with the Great Wall found in the Center for
Astrophysics Redshift Survey (Geller \& Huchra 1989), which has a linear
scale of roughly $240h^{-1}$Mpc. The size of the CfA Great Wall was
limited by the extent of that survey. In contrast, the SDSS is large
enough that a more extensive structure could have been found. 

The mapping method employed by Gott et al. is 
conformal, i.e., shapes of structures are
preserved locally, thus we can compare great walls and other
structures seen at different distance.
The x-axis is right ascension in degrees, while the vertical axis plots $y=\ln(r)$ where $r$ is the comoving coordinate
distance from Earth. In such a plot, the entire universe, from the
surface of the Earth to the epoch of the big bang, may be shown in one
plot.

\subsection{Cluster Catalogs}

Several algorithms have been applied to SDSS photometric and
spectroscopy data to identify clusters of galaxies (see, e.g.,
Nichol 2003 and references therein). Cluster catalogs have been
constructed from the SDSS imaging data alone using color criteria
(maxBCG: Annis et al.), adaptive matched filtering (AMF; Kim et
al. 2002), and a combination of these two methods (Bahcall et al. 2003).
Perhaps the most useful for the purpose of placing clusters in the
context of the cosmic web is the C4 method which
identifies clusters within the volume sampled by the main galaxy
spectroscopic sample. The C4 algorithm looks for density enhancements
in the 7-dimensional space of four colors (generated by comparison of
the five SDSS bands), right ascension, declination, and redshift. 800 clusters have been identified in the DR2 sample using this method (Miller et al. 2004).


\section{Galaxy Properties and Environment}
\label{sec:environ}

The environmental dependence of galaxy properties has
typically been described in terms of the well-known morphology-density
relation (Dressler 1980; Postman \& Geller 1984).
Study of the morphology-density relation using SDSS data (Goto et al. 2003) shows that this relation extends into the field population.
Recent analyses of
large spectroscopic samples of galaxies from the 2dFGRS and SDSS
reveal variation of galaxy properties, in particular dependence of
star formation rates, on density and/or clustercentric distance well
beyond the virial radii of clusters. Analyses of galaxies in voids
shows that this trend continues down to a small fraction of the mean
density.
Evidence suggests that the star formation rate depends not
only on the local density (on scales of 1 Mpc or so) but also on the
larger-scale environment. Here we highlight some recent results (also see Tanaka et al. 2004 in this proceedings) and describe some
possible mechanisms for this large-scale variation.

\subsection{Star Formation Rates as a Function of Clustercentric Distance}

Lewis et al. (2002) estimate star formation rates for galaxies in the
2dFGRS that lie near clusters. They find that the mean star formation
rates of galaxies, as well as the fraction of galaxies with high star
formation rates, increases with clustercentric distance, with a sharp
rise out to $\sim 2r/R_{virial}$. Analysis of SDSS galaxies near
clusters and groups found in SDSS (using the C4 method described
above) clearly shows a rise of the fraction of high star-forming
galaxies out to several virial radii.
Gomez et al. (2003) find similar results from analysis of the SDSS EDR sample and show that a ``break" appears in the density-SFR relation around a projected local density of $\sim 1 h_{75}^{-1}$Mpc.

\subsection{Star Formation Rate as a Function of Density}

Balogh et al. (2004) examine the variation of H$\alpha$ emission
equivalent width as a function of density for galaxies in
volume-limited samples of both 2dFGRS and SDSS ($M_r<-20.6$ for $h=0.7$). They
estimate galaxy density using both a projected density to the
fifth-nearest neighbor and a Gaussian kernel smoothing and find that
the fraction of galaxies with $W(H\alpha)>4\AA$ increases with
decreasing local density down to the lowest density that they can
reliably examine. Further, when they examine the fraction of high
$W(H\alpha)$ galaxies as a function of density estimated on both $1.1$
and $5.5$ Mpc scales, they find that, for fixed small-scale density,
high star formation fractions are larger for galaxies that lie in
large-scale underdensities. 
This result appears to disagree with Kauffmann et al. (2004), who find no discernable trend of star formation rate with large-scale environment. This discrepancy may result from the different range of densities probed (the lowest density bin probed by Kauffmann et al. includes over 30\% of the galaxies). In the next section we discuss what happens at truly low densities.

\subsection{Void Galaxy Properties}

To examine whether the trends identified near cluster continue all the
way into voids, we compare properties of void galaxies with those of
galaxies in denser regions (Rojas 2004; Rojas et al. 2004a,b; Hoyle et al. 2003, Hao et al. 2004).
What happens to star formation rates and related galaxy properties at a small fraction of the mean density?

To identify void galaxies, we measure the distance from each galaxy in the
flux-limited sample to the third nearest
neighbor, and require $d_3>7h^{-1}$Mpc for void galaxies in a volume-limited sample ($M_r<-19.87-5\log h$). 
This yields $1010$ void galaxies (6-8\% of the sample after
removing objects near the survey boundary) with $\delta\rho/\rho<-0.6$
on a $7h^{-1}$Mpc scale. Comparison of this criterion with results of
the voidfinder algorithm (Hoyle \& Vogeley 2002) show that it reliably
identifies galaxies that lie in large voids. At small redshift, we use
the nearby UZC and SSRS2 catalogs to map local voids, and identify an
additional $194$ fainter void galaxies.
This is the first sample of true void galaxies that is large enough to allow division into statistically-significant sub-samples.

Statistical analyses of the distribution functions of color,
surface brightness profile, and emission line equivalent widths, show
that void galaxies are, on average, bluer, more disklike, and have
higher specific (per unit stellar mass) star formation rates than
objects in denser regions (``wall'' galaxies hereafter). K-S tests
typically yield a probability $P<10^{-4}$ of the void and wall
galaxies being drawn from the same parent population. These same trends obtain when we compare void galaxies with the lowest-density 20\% of wall galaxies, thus these differences are apparent not only between ``cluster" and ``void" samples but also between true ``field" and ``void" populations.

Several analyses of SDSS photometry have shown that galaxies have a
nearly bimodal distribution of properties, either red and
bulge-dominated, or blue and disk-like
(Blanton et al. 2003; Baldry et al. 2004). 
To test the possibility that
void galaxies differ from wall galaxies simply because of a
demographic shift in the relative fractions of objects in these two
populations (fewer elliptical in low-density regions), we compare void and wall
galaxies in sub-samples with similar morphology and luminosity. We
find that void galaxies persist in having bluer average colors. Thus,
the morphology-density relation does not explain this
density-dependence.

In summary, we find that at fixed luminosity and surface brightness
profile, galaxies in voids have average colors that are bluer and
higher average specific star formation rates. These trends are generally
consistent with predictions of semi-analytic modelling of galaxy
formation (Benson et al. 2003).

\section{Discussion: Possible Causes of Large-Scale Environmental 
Dependence}
\label{sec:discuss}

What could cause these large-scale variations?  Is galaxy formation a
purely local process? Or does the large-scale environment contribute
to the fate of a galaxy? 
Environment can affect galaxy formation and evolution through a number
of factors that vary with density:
\begin{enumerate}
\item The dark halo mass function (shifts to lower masses at lower
  density)
\item The formation epoch of halo assembly
\item The rate of galaxy-galaxy interactions/mergers (including tidal
  effects and induced turbulence)
\item Temperature of the IGM
\end{enumerate}

The question of whether environment determines galaxy properties is
mostly physical, but partly semantic. The first point makes this
clear: large-scale environment affects the properties of the objects
found at different large-scale density through the mass function.
Likewise for the second point: for objects of fixed total mass today
but in different large-scale environments, the merging history of
halos is likely to be different. On the other hand, if we compare
galaxies of fixed total mass in regions of fixed small-scale density,
but in different large-scale environments, their properties are likely
to be similar (as found by Kauffmann et al. 2004), but perhaps not identical, as shown by the evidence cited above. 

\subsection{The Mass Function and Environment}

Following the Press-Schechter formalism, it is straightforward to predict how the
mass function of galaxies varies with large-scale
environment. In voids, the number
of small halos decreases in proportion to the lower overall matter
density, while the exponential cutoff at the high mass end shifts to
lower mass. This variation is consistent with the interior of voids acting like very low density universes (Goldberg \& Vogeley 2004). In Goldberg et al. (2004), we study this dependence,
compute the mass function for voids with varying $\delta_v$, estimate
the mass function of SDSS galaxies, and estimate $\delta_v$. 

Several paradoxical results relevant to the question of environmental
dependence on large-scales arise from this study. The growth factor
$D(z|\delta)$ in voids is larger at high redshift than in the
background universe, suggesting that void galaxy halos should be
relatively old. However, detailed analysis shows that, because the
mass function of void galaxies is quite steep at the relevant mass
scale, a small amount of growth at the present epoch yields a larger
fractional growth rate at the high mass end than in denser regions. In
other words, the few high mass galaxies in voids may have a high
fraction of recently accretion. This may explain why massive void
galaxies have relatively large star formation rates, while faint void
galaxies exhibit smaller differences from higher-density objects.

\subsection{Temperature, Environment, and Star Formation}

A consequence of the temperature dependence of the Jeans mass and
cooling rate of baryons is that we expect star formation to be more
efficient where the IGM temperature is lower. In trying to explain the
scale-dependence of bias in N-body/hydrodynamical simulations, Blanton
et al. (1999) examine how bias depends on gas temperature. They find
that, at fixed dark matter overdensity, the overdensity of baryonic
matter that can form stars is larger in region of cooler gas
temperature. 

Because the gas temperature reflects fluctuations in density on scales
much larger than the galaxy mass scale, star formation may be
modulated by the large-scale density field. When the gas is
virialized, its temperature reflects the local gravitational potential. Poisson's
equation implies that $\tilde{\delta}_T(k)\propto k^{-2}\tilde{\delta}(k)$ (Blanton et al. 1999). In other words, the extra two powers of $k$ cause fluctuations in temperature to depend on much larger wavelength fluctuations in density than the density field itself. Thus, the large-scale environment strongly influences the IGM temperature, through which it may strongly affect the efficiency of star formation.


\begin{acknowledgments}

M.S.V. acknowledges support from NSF grant
AST-0071201.

Funding for the creation and distribution of the SDSS Archive has been
provided by the Alfred P. Sloan Foundation, the Participating
Institutions, the National Aeronautics and Space Administration, the
National Science Foundation, the U.S. Department of Energy, the
Japanese Monbukagakusho, and the Max Planck Society. The SDSS Web site
is http://www.sdss.org/.
                      
The SDSS is managed by the Astrophysical Research Consortium (ARC) for
the Participating Institutions. The Participating Institutions are The
University of Chicago, Fermilab, the Institute for Advanced Study, the
Japan Participation Group, The Johns Hopkins University, Los Alamos
National Laboratory, the Max-Planck-Institute for Astronomy (MPIA),
the Max-Planck-Institute for Astrophysics (MPA), New Mexico State
University, University of Pittsburgh, Princeton University, the United
States Naval Observatory, and the University of Washington.
\end{acknowledgments}

 \end{document}